# INTERFERENCE IN OFDM SYSTEMS AND NETWORKS

**Tomáš Páleník, Viktor Szitkey**

*Abstract:*

*In this paper we present an overview of various kinds of interference, that arise in the Orthogonal Frequency-Domain Multiplexing (OFDM)-based digital communications systems at the physical layer. Inter-symbol, inter-block, inter-carrier interference types are described in detail, valid for any OFDM transmission, along with Inter-cell interference specific to cellular networks. A survey of various communication disruptions techniques is presented – primarily focusing on intentional interference – jamming. Furthermore, we present a survey of modulation techniques that expand on the OFDM and may be considered viable candidates for modulation in the future 6G cellular networks.*

*Keywords:*

*OFDM, interference, jamming, 6G modulation candidates.*

## Introduction

Orthogonal Frequency Division Multiplexing (OFDM) is the most commonly used wide-band modulation technique, present in nearly all modern high bandwidth communication systems and standards. Employed reliably in Wi-Fi networks, it serves reliably at the physical layer as modulation in IEEE 802.11a, g, n, ac, ax, as well as the latest iteration – IEEE802.11be [1], also known as Wi-Fi 7. Given its high spectral effectivity and resilience to multipath fading occurring in terrestrial transmissions, OFDM has been the modulation of choice also in the context of cellular networks for some time now: the time-proven 4G Long Term Evolution (LTE) [2] as well as the currently deployed 5G networks [3] are both based on the OFDM modulation. OFDM serves also in the wired digital subscriber lines standard DOCSIS [4], and even in the proprietary Starlink LEO satellite constellation (The specification is not publicly available, but measurements are [5]). Furthermore, various modulation schemes are currently evolving from the OFDM, and are subject of intense theoretical research, evaluation and debate, with several potential candidate technologies contemplated for utilization in the future 3GPP 6G standards releases on the horizon 2030.

However, OFDM is also susceptible to various forms of interference, occurring naturally or as a result of intentional disruption effort: jamming, where malicious actors disrupt communication by emitting interference signals. To counter such threats, various techniques of OFDM modulation fortifications are emerging. The analysis of intentional disruption techniques is an important input for future standardization. The 3GPP standard 5G New Radio (NR) [4] is currently being deployed on a global scale and can be viewed as a direct evolution of the OFDM-based transmission defined for 4G.

The primary security measure in wireless networks is the encryption of the over-the-air messages, providing protection from eavesdropping and, as described in later sections, also some indirect attack resistance.

Since the time of the failed and completely broken Wired Equivalent Privacy (WEP) attempt, encryption is heavily employed in Wi-Fi. The modern Wi-Fi Protected Access (WPA) [6] security framework is currently in its third generation with Perfect Forward Secrecy (PFS) as an integral feature [7]. In the context of cellular networks, serious effort to encrypt the control information sent out by the base station has been put already into the 4G standard, while some vital pieces of information, that can be used by the attacker to disrupt network operation, are transmitted out in plaintext. On top of that, some control messages are not protected by a cryptographic Message Integrity Check (MIC), giving a potential attacker the opportunity to falsify them. Most of these weak points have been addressed in the 5G NR specifications with one important caveat: to accelerate the deployment of 5G, two potential modes of operation have been standardized: the Standalone Application (SA) [8], also called the true 5G, is the desired variety in which the 5G base station, known as the gNodeB, is connected to the 5G core network. The Non-Standalone Application (NSA) is a backward-compatibility measure designed as an intermediate step to integrate newly deployed gNodeB stations into existing 4G network: The 5G gNodeB provides fast data-frames transmission to the UE, while the control messages are all handled by the previous-generation cooperating eNodeB [9]. Most of the currently deployed 5G networks in the European union operate in a NSA mode, with full SA to be implemented as soon as possible. Based on this insight, when analyzing the vulnerability of a 5G NSA network, many of the weaknesses identified for 4G networks are still valid.

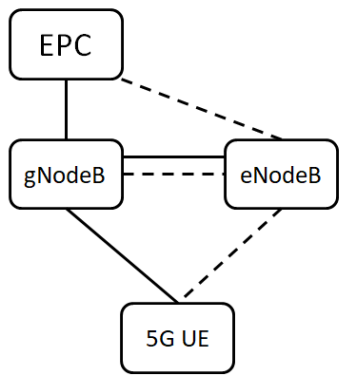

Fig.1. NSA 5G network operation, option 3x [9]: The 5G gNodeB provides fast data-frames transmission to the UE (solid line), while the control messages (dashed line) are all handled by the previous-generation cooperating eNodeB.

The paper is organized as follows: the next section gives a brief recap of the principles of wireless OFDM transmission in the multipath fading channel environment of cellular networks. The second section describes various forms of "naturally" occurring interference – interference inherent to the OFDM design. The third section then provides a survey of recent literature – the current state of published research on topics of unintentional interference. Intentional interference also known as jamming, is surveyed in the fourth section. The fifth section then provides an overview of the potential OFDM modifications – evolution steps, that may be considered as candidates for modulation techniques in the future 6G specifications. The last section then concludes the paper.

# 1 OFDM Principles

The basic principle of OFDM is the demultiplexing of a high-rate data stream to many low-rate streams in the transmitter, with each of the substreams modulated – assigned a prototype in digital domain (signal space) and placed on a separate subcarrier frequency. The $N$ subcarriers are spaced uniformly in the frequency domain with distance $\Delta f = 1/T$, where $T$ is the OFDM symbol duration. This subcarrier spacing provides for the orthogonality of the subcarriers as illustrated in (Fig.2): The center frequency of each subcarrier is simultaneously exactly the zero-crossing point of the frequency responses of all the other subcarriers, making the set of discrete frequencies $f_i \in \{k\Delta f + f_0\}$, $k = \{0, 1, \ldots N-1\}$ points of zero interference. The set of $N$ orthogonally spaced subchannels is then multiplexed together to form the resulting, potentially complex, OFDM prototype. Inverse Discreet Fourier Transform (IDFT) is utilized for practical implementation of this process along with converting the frequency domain digital prototype samples to time domain.

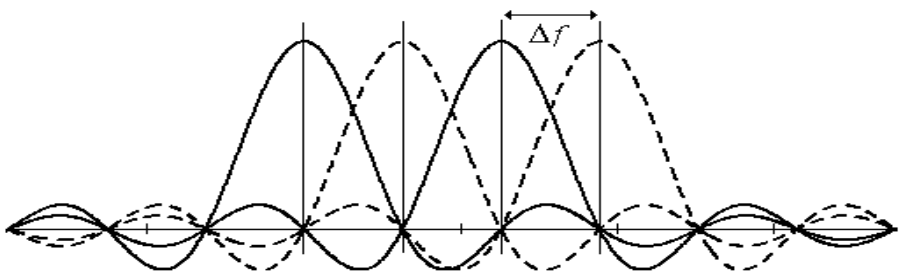

Fig.2. Orthogonal subcarrier spacing in frequency domain [10].

A unitary version of the IDFT given by (1) may be chosen to preserve the signal energy.

$$x_n = \frac{1}{\sqrt{N}} \sum_{n=0}^{N-1} X_k \cdot e^{\frac{j2\pi}{N}nk} \quad n,k=0,\ldots,N-1 \qquad (1)$$

After the conversion to time domain, the block of samples of the OFDM symbol is prepended by its own subblock containing some small fraction (1/32 to 1/4) [11] of the symbol samples. This repetition of some of the samples is called the Cyclic Prefix (CP) Insertion (CPI) and together with an inverse operation in the receiver – the CP Removal (CPR) it serves the important function of facilitating a multipath channel distortion equalization. The effect of the CPI on the data samples is shown in (Fig.3) with the CP length expressed in time denoted as $T_g$, the useful symbol time $T_u$, and the final transmitted OFDM symbol duration $T = T_g + T_u$. Alternatively the CP length may be expressed in the number of samples: If the number of useful samples is $N$, then the total number of the symbol with CP may be denoted $N_c$.

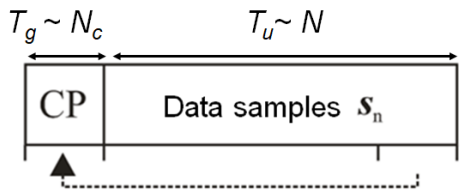

Fig.3. The OFDM symbol (block of samples) prolonged by cyclic prefix insertion.

In a wireless transmission, the primary effects negatively effecting the signal are reflection, refraction and scattering [10]. As shown in (Fig.4), several delayed and attenuated copies of the transmitter signal arrive at the receiver antenna where they combine – usually in a destructive way.

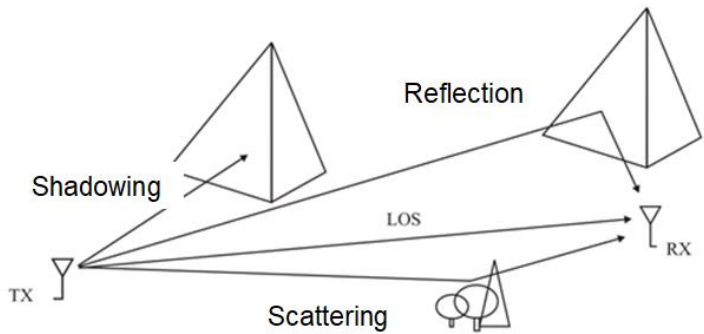

Fig.4. Multipath propagation of a wireless signal in a terrestrial channel [10].

The digital data transmission over a multipath channel may be simply modelled by a convolution with channel impulse response h in the time domain given by (2), and addition of an Additive White Gaussian Noise (AWGN) with power $\sigma^2$.

$$r_n = \sum_{m=1}^{v} h_m \times t_{n-m+1}, \quad n = 1,...,N_c + v - 1 \qquad (2)$$

A simplified OFDM system is shown in (Fig.5) together with the multipath channel model. While the Analog/Digital and Digital/Analog converter circuits are necessary, the focus of most of the signal processing described in literature lies in the digital domain. The ECC block presents the Error Control Coding – adding redundancy to data in order to protect the data against errors introduced by the channel, while the corresponding A-Posteriori Probability (APP) block in the receiver denotes a soft-input ECC decoder based on posterior probabilities. The Constellation Mapping (CM) block performs the signal space mapping described above and the Discrete Fourier Transform (DFT) block in the receiver transforms the received OFDM block back to frequency domain. The noisy samples then enter the Frequency Domain Equalization (FDE), after which the projections to the signal space are used as input to the ECC decoder.

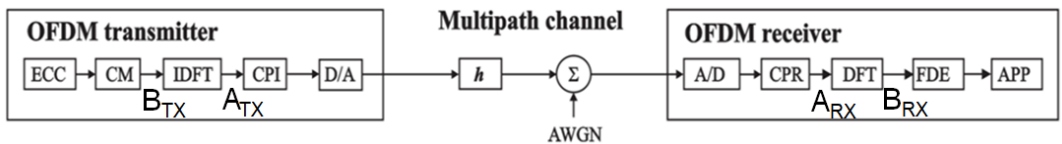

Fig.5. OFDM transmission model with multipath channel [12].

The cyclic prefix insertion in the transmitter, and its removal in the receiver converts the linear convolution that actually takes place in the channel (2) to a cyclic convolution defined by (3) effectively observable between the interfaces $A_{TX}$ and $A_{RX}$ shown in (Fig.5).

$$r_n = \sum_{m=1}^{\nu} h_m \times s_{[(n-m) \bmod N]+1}, \quad n = 1,...,N \tag{3}$$

By utilizing the DFT properties, the cyclic convolution of the discrete-time domain vectors translates to simple per-component multiplication of their Fourier transforms in the discrete-frequency digital domain, so the cascade of blocks shown in (Fig.5) between interfaces $B_{TX}$ and $B_{RX}$ can be expressed as a set of $N$ independent simple per-sample multiplications. This may be written by (4) where the $k$ is the index in the discrete-frequency domain $H(k)$ is the channel frequency response, the $S(k)$ is the transmitted OFDM symbol – a block of complex numbers of length $N$, and $R(k)$ is the received block of distorted and noisy samples at the input of the FDE block in the receiver.

$$R(k) = H(k) \cdot S(k), \quad k = 0,...,N-1 \tag{4}$$

Given a channel frequency response estimates $H(k)$, the equalization performed in the frequency domain described equally simply by per-sample division as shown in (5):

$$\hat{S}(k) = \frac{R(k)}{\hat{H}(k)}, \quad k = 0,...,N-1 \tag{5}$$

The computational simplicity of the frequency domain equalization of the multipath channel distortion is the primary reason for the success of the OFDM modulation, allowing its ubiquitous presence in almost all modern wideband digital communications systems.

## 2 Interference in OFDM Transmission

The simplified model of an OFDM transmission presented in the previous section omits the various conditions and interference types that negatively affect an OFDM system. Further blocks need to be added to the simplified system model shown in (Fig.5), in order to mitigate various types of interference and other negative channel effects. Some of them explained in later sections. Let us now summarize the various interference types that negatively affect OFDM transmission.

**Inter-Symbol and Inter-Block Interference**

One of the effects of the linear convolution of the transmitted OFDM symbol $S(n)$ by the channel impulse response $H(n)$ is the prolonging of the transmitted block: As defined in (2): if the length of the transmitted block $S(n)$ is $N_c$ and the length of $H(n)$ is $\nu$, then the length of the resulting received block is $N_c + \nu - 1$. As the OFDM symbols are transmitted serially, some samples of the received symbol begin to overlap with the next symbol. This effect, when expressed in discrete-time, when the OFDM symbol is represented by a vector of complex numbers, is called the Inter-Block Interference (IBI). When viewing the OFDM in the continuous time as a waveform, this effect is called Inter-Symbol Interference (ISI). As described in the next section, this is somewhat confusing, and a more precise term: Inter-OFDM-Symbol Interference (IOSI) would be more suitable. The general practice in literature is the usage of the term ISI, while also using the terms ISI and IBI interchangeably.

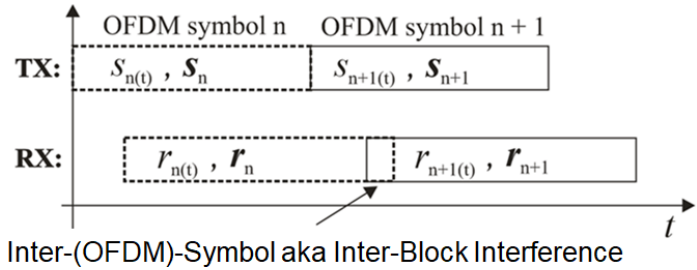

Fig.6. Inter-(OFDM)-Symbol and Inter-Block Interference
as a consequence of linear channel convolution [12].

**Inter-Symbol Interference**

As described before, the CP prefix insertion and removal in OFDM serves the important function of removal of the Inter-Block Interference. This is illustrated in more detail in (Fig.7), that shows that the transmitted OFDM symbols with redundant CP are indeed affected by the IBI, but as long as the length of the CP is longer than the channel maximum excess delay, only the CP is affected, and the useful part of the OFDM symbol remains IBI-free. The condition of CP length being longer than the channel maximum excess delay is fundamental for successful equalization an OFDM the receiver.

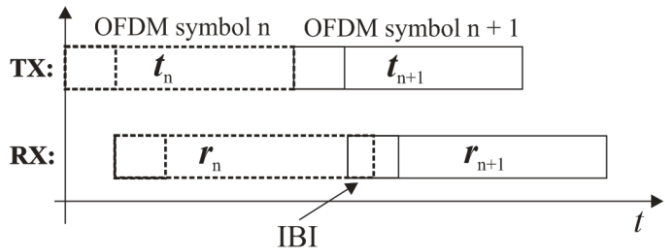

Fig.7. Transmission of OFDM symbols with redundant Cyclic Prefix:
IBI affects only the CP samples, which are discarded in the receiver [12].

After the conversion of the linear channel convolution to circular convolution defined in (3), the IBI goes away, but each of the received samples within one OFDM block is still a linear combination of several transmitted samples. Such linear convolution may be modelled by a simple matrix multiplication with a circulant channel convolution matrix $\mathbf{H_c}$:

$$\mathbf{H_c} = \begin{bmatrix} h_1 & 0 & \ldots & h_v & \ldots & h_2 \\ h_2 & h_1 & 0 & 0 & \ddots & \vdots \\ \vdots & h_2 & h_1 & \ddots & 0 & h_v \\ h_v & \vdots & h_2 & \ddots & 0 & 0 \\ 0 & \ddots & \vdots & \ddots & h_1 & 0 \\ \vdots & 0 & h_v & \ldots & h_2 & h_1 \end{bmatrix} \quad (6)$$

This effect is again known as Inter-Symbol Interference (ISI) and is qualitatively different from the IOSI defined in the previous section in that all the received samples originate with the same transmitted OFDM symbol/block. In literature the term ISI is used interchangeably in both situations and the reader must be aware of the context in order to understand which effect is actually being described.

**Inter-Carrier Interference**

The interference type with the most disastrous effect on the transmissions Bit Error Ratio (BER) is without doubt the Inter Carrier Interference (ICI). Again, the nomenclature is imprecise, since what is actually going on in this case is the loss of orthogonality between subcarriers, belonging to the same multiplexed set of substreams of one transmission. The more precise term should therefore be Inter-SubCarrier Interference (ISCI), but ICI is the dominant term in literature. As illustrated in (Fig.2), in an ideal OFDM system subcarriers are orthogonal to each other, meaning their waveforms do not interfere at the select discrete subcarrier frequencies, even as they otherwise completely overlap in the frequency domain. In a practical system, the subcarrier may be slightly shifted for various reasons: for instance the local oscillator in the receiver analog frontend may have a slightly different frequency than the oscillator in the transmitter. This is denoted as frequency offset an illustrated in (Fig.8), where all the subcarrier frequencies are shifted by some small $\Delta f$. As illustrated, the spectrum of one subcarrier is now perturbed by the sum of the spectra of many different subcarriers at this frequency point, not just the directly adjacent ones: each of the discrete frequency points in (Fig.8) consists of the sum of 4 different values where the largest value represents the desired signal, and the sum of all the others represents harmful interference. Even a small frequency shifts cause a substantial accumulation in this sum and a subsequent drop in the perceived Signal to Noise Ratio (SNR), dramatically increasing the error probability. In wideband systems, the total channel width is in the order of tens of MHz (the typical LTE channel width is 10 MHz, while the classic Wi-Fi channel is 20 MHz wide). Much higher values are permitted in more recent version of these standard with IEEE 802.11 be specifying channels up to 320 MHz wide [1]. The problem is that the subcarrier spacing is much tighter than that, e.g. in the LTE, it is only 15 kHz, which is, comparing to the channel bandwidth, several orders of magnitude smaller. For such a small subcarrier spacing, even a few kHz of frequency offset, when unmitigated, represents a disastrous ICI.

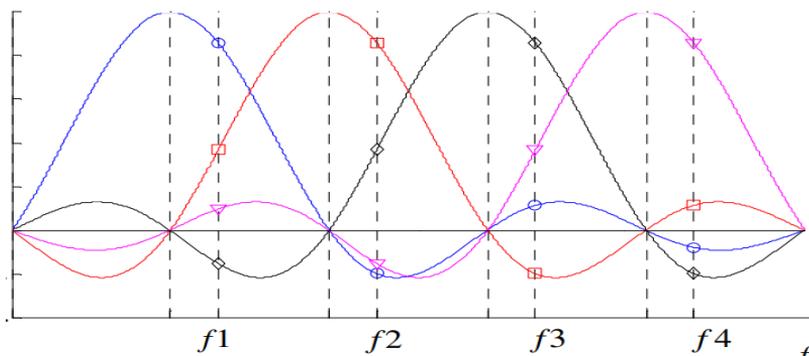

Fig.8. The loss of orthogonality between subcarriers because of RX frequency offset [13].

Even though the local oscillator offset may be successfully eliminated by a Phase-Locked Loop (PLL) circuitry in the receiver, other effect, such as the Doppler shift or amplifier nonlinearities may introduce further spurious frequency components together with the another complication: the frequency offset is not constant over the systems bandwidth. The ICI also arises if the fundamental condition for FDE: the CP length exceeding the channel excess delay, is not satisfied.

**Inter-Cell and Intra-Cell Interference**

Mobile communications networks are called cellular networks because of the fact that the area of interest is divided into a grid of non-overlapping cells – each with a base station in the middle, covering a carefully selected subarea. The cells are large in diameter in case of sparsely populated rural areas, and very small in dense urban centers. A typical diagram of part of cellular network is shown in (Fig.9). While in reality the actual cells may take on complex shapes, reflecting the actual topology – the shapes and sizes of the buildings in case of a urban setting, the cells are usually drawn as a hexagonal grid for simplicity. As shown further in (Fig.9) using this approximation each cell borders on 6 other cells. While the signal reception is good near the cell center, as the UE moves between cells, it must spend some time in the cell border area, where the signals from two, or even three base station overlap. The UE is usually connected to only one serving base station, so the signal from neighboring BSs presents an interference – the Inter-cell interference. In contrast to the inter-cell interference the intra-cell interference is used as an umbrella term for various other types of interference occurring within one cell as described in previous sections.

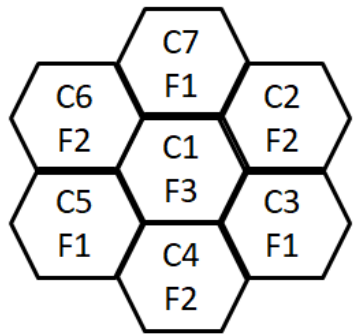

Fig.9. The basic structure of a cellular network with Frequency Reuse factor 3: using F1, F2, F3.

Several methods of dealing with this type of interference are applied in various cellular networks generations with one of the oldest – the Frequency Reuse Factor 3 [14] shown in (Fig.9): the available bandwidth is divided in three different subbands, while each of the cell uses one of the subbands with no two adjacent cells using the same subband. In this way, the inter-cell interference is successfully eliminated for the price of reducing the available frequency bandwidth by a factor of 3. Given the astronomical prices the network operators pay for the licensed bands, this price is too high to pay and one of the specifics of the OFDM is the ability to implement interference coordination strategies like Coordinated Multi-Point (CoMP) in a Single Frequency Network (SFN) [15] where even the neighboring cells can use the same frequency, but the base stations synchronize their OFDM transmission in such a way, that the UE receives several copies of the same OFDM symbol, originating from different base stations. These copies then don't interfere, they combine constructively, and if the time difference between the first and the last received copy doesn't exceed the Cyclic Prefix length, the OFDM receiver is in theory able to decode these signals as a standard multipath channel using the FDE described before. Furthermore, this coordinated transmission, denoted depending on the context as spatial- or antenna- diversity, may actually turn the interference into a benefit to the transmission: if the signal from one BS is temporarily overshadowed by some obstacle, the signals from neighboring cells may happen to coincide with a better path with less attenuation.

**Self-, Cross Link- and Adjacent Channel- Interference**

The problem of self-interference arises in situations when the collocated transmitter and receiver both transmit and receive at the same time. This may occur in various scenarios but the most likely is the operation of a base station in cellular networks. In older Frequency Division Duplex (FDD) specifications the downlink (DL) and uplink (UL) directions (to and from the UE) are separated by using different frequency bands, while in more recent 5G NR Phase 2 systems the Time Division Duplex (TDD) becomes dominant (depending on the exact band used). In TDD the UL and DL directions use both the same frequency band, but not simultaneously – at any given time transmission occurs in only one of the directions. The most recent advancements, contemplated for deployment in the 5G-Advanced, present a new scheme for a full duplex operation: the Sub-band non-overlapping Full Duplex (SBFD) [16]. In this duplex mode, it is possible to receive and transmit at the same time within the same band, but the band is further partitioned to subbands. (Fig.10) shows a standard OFDM DL resource grid: a grid of time- and frequency- slots, as used in the LTE along with some terminology: the Resource Element (RE) is the smallest unit and consists of one subcarrier with duration of one OFDM symbol. The REs are grouped into larger groups – rectangular Resource Blocks (RB) to be allocated for mapping of user data. Similar structure, differing only in exact parameters values (aka numerology) is valid for 5G. In the SBFD context, dynamically created subbands may be allocated within this grid and may be used for UL transmission.

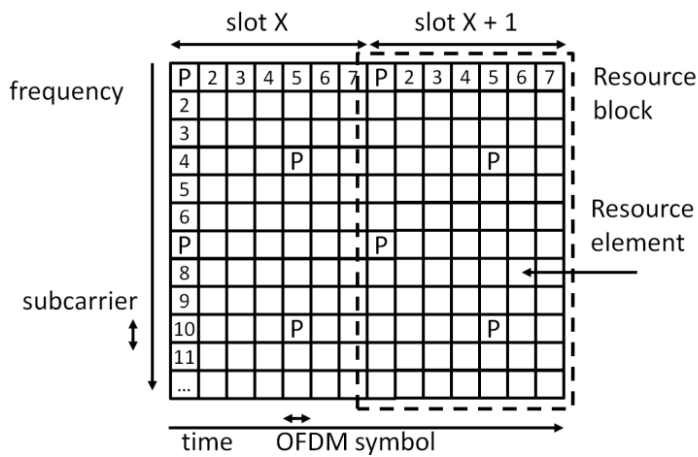

Fig.10. OFDM resource grid in LTE.

Selected subbands may be then dynamically allocated for UL transmission, while the rest of the subbands carry downlink data. The method is then a combination of FDD and TDD, with the advantage of much finer granularity for subband selection and increased flexibility for UL and DL resources allocation [17]. However, it also (re-)introduces another interference types: The *Self-Interference* is the interference of the transmitter, whose output radiated field interferes with the received signals from the opposite side of the wireless link. It is usually a problem occurring at the base station, since it uses a large output power and signal arriving from a connected UE is much weaker. In a similar fashion, the *Cross-Link Interference* (CLI) denotes such potential interference between DL and UL signals from different UEs or coming from a different cell.

An in principle somewhat similar type of interference is the *Adjacent Channel Interference* (ACI) described in [18] as an interference that arises from the fact that one of the inherent drawbacks of OFDM is the high Out-Of-Band (OOB) radiation.

This stems from the construction of the OFDM prototype by setting the complex amplitudes of a continuous set of subcarriers as shown on (Fig.11), which can be understood as using a rectangular window, which causes a significant fraction of energy overflowing from modulated subcarriers to adjacent frequency bins.

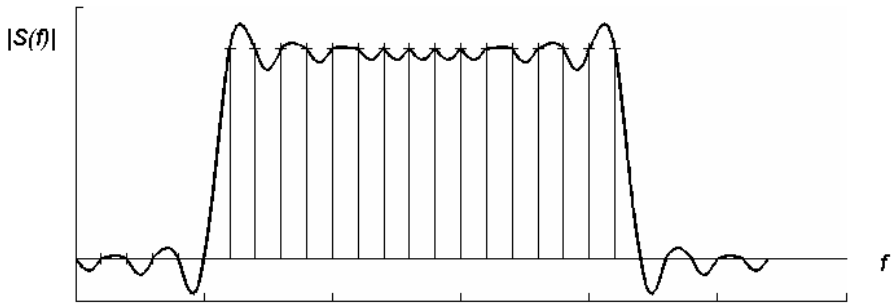

Fig.11. Setting the values of OFDM subcarriers in frequency domain
may be viewed as using a rectangular window.

A common means of dealing with this interference is to set some fraction of the subcarriers at the edges of the band to zero as a guard band. On the other hand, this approach introduces waste of precious frequency bandwidth, so alternative method of subcarrier-wise filtering and Filter-Bank MultiCarrier (FBMC) implementation may be used [19].

## 3 OFDM Interference & Mitigation Survey

Since the Inter-Carrier Interference is the most destructive form of interference, it has been the subject of intense research over the years with some limited success achieved in the form various ICI-Self Cancellation (ICISC) designs. All based on the observation, that adjacent subcarriers, if they are modulated with opposite polarity symbols, can be used in the receiver to cancel out each other's interference contribution. This has been described in literature and expanded on in the form of introduction of Polynomial Cancellation Coding (PCC) [20], Symmetrical Cancellation Coding (SCC) [21] and Conjugate Cancellation (CC) [22]. All of these ICI cancellation schemes share a significant drawback: they may all be understood as some form of repetition coding with low rate around $R = 1/2$. Therefore, they effectively reduce the useful data rate (and OFDM spectral effectivity) to a half. [23] represents a more recent significant evolution of the interference cancellation by introducing two novel ICI cancellation schemes: one based on a especially designed data scrambling algorithm, other based on interleaving the complex subcarrier amplitudes in the frequency domain prior to the IFFT. Both proposed schemes do not introduce any extra redundancy, while providing some ICI cancellation capabilities presenting an interesting compromise between ICI cancelling and added complexity.

Similarly [24] describes a method for mitigation of the ICI in the context of OFDM communication with passive RF tags. These simple devices have no power source and rely on reflecting and altering the received OFDM signal according to the information to be transmitted. This is known as backscattering and has been traditionally used with simpler modulations such as ON-OFF keying. The novel usage of backscattering of a complex OFDM transmission has been suggested only recently in [25] with the limitations arising from the spectral shift occurring because of the backscattering process, causing ICI. [25] then proposes an interference cancellation scheme that utilizes unused OFDM subcarriers for generating of the cancellation signal for ICI suppression.

[26] describes an ICI mitigation scheme in the context of MIMO-equipped OFDM systems employing advanced signal processing in the form of Space-Time Line Code (STLC) or Space-Time Frequency Code (STFC) [27], that enable the encoding of transmitted symbols using the Channel State Information (CSI) in the transmitter, while decoding using a simple linear combination with only partial CSI in the receiver, achieving full rate together with full diversity. The weakness of STLC/SFLC-OFDM systems lies in the sensitivity to ICI caused by the Doppler shift in high-mobility scenarios. [27] then reuses the principle of the ICI- Self Cancellation defined in [28] that manipulates the symbols assigned to adjacent subcarriers in order to cancel out the ICI. Specifically, the data bits are modulated in such a way, that adjacent subcarriers carry symbols with opposing values.

A complementary scheme for ICI mitigation in high-mobility setting is described in [29] again in the context of MIMO-OFDM systems, while deriving a custom method of signal detection based on the Expectation Propagation (EP) method described in [30]. The proposed new method based on Bayesian estimation [31] performs data estimation assuming the perfect Channel State Information (CSI) is available in the receiver, and also suggest a more practical possibility of using only estimated CSI pointing out to [32]. The simulation results provided indicate good error performance together with a relatively low complexity implementation.

In the context of Visible Light Communication (VLC) a modified OFDM scheme must be used, with the necessary constraint, that the optically transmitted signals must have nonnegative amplitudes. This gave rise to OFDM variants such as Asymmetrically Clipped OFDM (ACO-OFDM) [33] and Direct Current biased Optical OFDM (DCO-OFDM) [34]. Interference arises due to the clipping noise, and a pre-distortion is suggested and described in [35] to eliminate it. The novel pre-distorted scheme is then termed Pre-Distorted or PADO-OFDM.

The Inter-Cell Interference, as a major problem in cellular networks, attracted significant research interest in the past [36] as well as in recent years [37][38]. One approach is to coordinate transmissions between base stations of neighboring cells to perform one or a combination of the following techniques: Interference Alignment (IA) [38], or coordinated beamforming [39]. IA is in principle based on orthogonalization of the interfering signals: if we can, by some form of signal processing, force the interfering signal to be orthogonal to the useful transmission, then they actually do not interfere at all. As shown in [36, 38], this can be done to some extent, provided we sacrifice some of our signal space dimensions. In the simplest example by using only the real axis while the interfering transfer occupies only the imaginary axis of the IQ plane. In [40] Interference Alignment is discussed in detail for the context of $K$ interfering users and it is shown, that the sum-spectral efficiency can be made to scale linearly with $K/2$, for which a modified transmission scheme termed the Interference-Free OFDM (IF-OFDM) is derived.

The coordinated beamforming is based in the ability of antenna array on the base station to direct the EM field in a narrow beam specifically pointed at the UE for which the data is heading. Then, if the neighboring cells are cooperating, such spatial beams may be optimized to minimize potential ICI. These mechanisms are often referred to as Inter-Cell Interference Coordination (ICIC) [41]. A complementary approach is to attempt to eliminate the ICI using no coordination, where the interference parameters are to be detected by the receiver alone. This is known as Blind Interference Parameter Detection (BIPD) [42][43] and usually comes with prohibitively complex receiver signal processing. A more recent approach is a Semi-BIPD based on utilization of received Demodulation-Reference Signals (DM-RS) sequence is described in [44].

When dealing with interference, an important research area deals with intentional OFDM disruption. This may be further divided into spoofing, sniffing and jamming, with the latter in focus of this survey. Jamming can be understood as intentional artificially created interference and there are many different techniques described in literature. Perhaps one of the simplest being the Narrow Band Jamming (NBJ), synonymous to Narrow Band Interference (NBI), or in some context denoted as Single-Tone Jamming/Interference. The principle is always the same: a narrow-band signal interfering with only a selected subset of the subcarriers. The Single-Tone Interference is analyzed in [45] both theoretically and by means of MATLAB simulation. Two scenarios are described: first with the jamming tone aligned with the target OFDM grid so that its frequency exactly matches one of the subcarriers, the second with a non-aligned interferer where the energy of the interfering signal flows over to several subcarriers. In [46] a somewhat similar scenario is presented with focus shifted to Chirp Spread Spectrum (CSS) system interfering with target OFDM transmission. The CSS modulation, used in the prominent IoT communication protocol LoRa, is based on time-variation of the signals frequency over a wide bandwidth. However, at each time instant, the CSS signal is a very narrowband waveform close to a single tone. [46] utilizes a special form of Fourier transform – the Fractional Fourier Transform (FRFT) as a tool necessary to properly analyze such CSS signal and then uses the resulting modified spectrum to derive an interference suppression algorithm. In [47] interference and jamming is analyzed in the context of Low-Earth Orbit (LEO) OFDM transmission: a variance-based Forward Consecutive Mean Excision (FCME) is utilized for detection of a narrowband interferer within the OFDM wideband channel and fast identification of the targeted subcarrier, while an overlapping windowed DFT frequency domain notch-based interference suppression is evaluated. In [48] an intereference cancellation technique based on neural networks is used for narrowband interference suppression. Another recent take on the NBJ problem (denoted somewhat differently as Spectral Concentrated Interference - SCI) is presented in [49], where a preliminary design of a combined OFDM-Spread Spectrum system is proposed.

[50] proposes a powerful jamming technique for disrupting OFDM, similar to the Repeat-Back Jamming technique [51], known for spread-spectrum systems. The proposed method disrupts the cyclic prefix of the OFDM symbols, which is essential for mitigating the interference caused by multipath fading. As long as the CP is longer than the maximum channel excess delay, the linear channel convolution can be converted to circular convolution for computationally feasible Frequency-Domain Equalization (FDE) described in section two. The key idea presented in [50] is to invalidate this condition by introducing a signal path with a delay exceeding the CP length. The jammer achieves this by capturing the transmitted OFDM signal and retransmitting it with a deliberate excessive delay. This creates a situation where the combined original and jamming signals mimic a multipath channel with an extended delay that exceeds the CP's length. Consequently, both IBI and consequent ICI cannot be fully suppressed by the receiver, disrupting the OFDM transmission. Since the jamming signal is simply a delayed version of the transmitted signal, the attack is difficult to detect. Unlike other jamming methods, this approach requires no synchronization with the OFDM transmission or knowledge of its detailed frame structure. On the other hand, it requires a powerful and ultra-fast signal processing on the jammer's side, since the FDE must first be performed and the transmitted block reconstructed by the jammer, sometimes in order of nanoseconds.

[52] then presents a potential mitigation technique based on attempting of a reconstruction of the transmitted OFDM symbol based on the decision feedback, where the one-tap FDE which is the heart of OFDM, is replaced by a more complex multi-tap FDE. This can reduce the error-floor otherwise introduced by the ICI that results as a consequence of exceeding CP length for the price of substantial increase of the receiver algorithm complexity. More recently [53] then presents an evolutional step in the form of Reduced Complexity Interference Cancellation (RCIC) with shorter detection delay, and reduced complexity with a compromise in resulting error performance.

Potentially the most effective class of jamming methods is the so called intelligent or smart jamming. Although these methods vary in details, they are based on the principle of understanding the organization of the OFDM resource grid as shown in (Fig.10), showing the allocation of data subcarriers along with pilot ones. What the structure indicates is that when it comes to intentional disruption, the subcarriers are not created equal: An OFDM receiver must estimate the channel frequency response $H(k)$ before it can perform the frequency-domain equalization as described by eq. (5). To enable this, OFDM systems transmit pilots, or reference symbols (RS), on pre-defined subcarriers alongside the data-carrying symbols. In 3GPP specifications, these reference symbols are generated at the PHY layer and referred to as the Cell-specific Reference signal (CRS). The CRS occupy approximately 14% of the resource elements and their exact positions and values are derived from the Cell ID [54]. Similarly, the synchronization signals represent only a small portion of the transmitted OFDM grid, and are therefore a potential target for jamming attack. [55] provides an analysis LTE transmission jamming based on the spoofing of the synchronization signals and gives simulational results confirming the effectivnes of this approach.

As described in [56], jamming of the pilot signals is a very effective way of disrupting OFDM transmission, provided the jammer can first identify the synchronization signals sent out by the base station in regular intervals. This method may be denoted as synchronous jamming, based on the fact that the jammer must synchronize to the network. As further shown in [54], perfect synchronization is not even necessary if the OFDM symbol duration is sufficiently long, and even an asynchronous disruption is feasible if the jammer transmits noise on all potential pilot subcarrier positions. Another highly effective disruption method is identified by [57] based on the knowledge of the 3GPP frame and superframe structure – the disruption of the Physical Broadcast Channel (PBCH) that caries the vital Master Information Block (MIB) header. Since the MIB provides the UE with important PHY-layer control parameters, a proposed intelligent jammer PBCH-IJ will target only the resource elements carrying the MIB to disrupt its decoding by the UE. As described in [57] this method represents a serious Denial of Service (DoS) attack.

## 4 OFDM Modifications and Candidates for 6G Modulation Scheme

Pre-standardization research activities are currently ongoing that aim at providing an answer to the question, what will the sixth generation (6G) of mobile communications be able to achieve and provide. The vision commonly held is that 6G will provide a technological platform enabling a unification of physical, digital and human worlds. It is expected to feature use of AI, machine learning, incorporation of integrated sensing and communication and more.

Already existing and new use cases will dictate requirements like extreme peak throughputs in orders of 100 Gbps, ultra-low latencies down to 0.1 ms not to mention the need for increased reliability, efficiency, security, etc... [58]. These requirements render existing modulation formats in their current form inadequate, prompting development of new modulation.

The choice of modulation format or waveform triggered discussions in earlier generations, with OFDM being used in both 4G and 5G. Depending on scenario OFDM is foreseen to be also used in 6G as no other waveform has shown substantial gain over OFDM not to mention the desire for efficient sharing of spectrum with the previous generation [59]. With that in mind is worth to mention few waveforms that are being considered as candidates for 6G networks.

Filtered OFDM (F-OFDM) is a variant of OFDM that introduces pulse shaping to reduce out-of-band emissions. It employs a filter on the subcarriers to reduce the side lobes, which helps to improve spectral efficiency and minimize interference with adjacent frequency bands [60].

Filter Bank MultiCarrier (FBMC) compared to OFDM, differs in three major ways. Firstly, it uses OQAM mapping instead of QAM mapping. With appropriate filtering this approach results in reduction of ICI. Secondly after IFFT process Poly Phase Network (PPN) filtering is applied, reducing ISI and ICI. Thirdly, no cyclic prefix is used because the best frequency and time localization through filtering and OQAM modulation process [61].

Universal Filtered Multi-Carrier (UFMC) can be seen as a generalization of F-OFDM and FBMC. In contrast to filtering entire band in F-OFDM and individual subcarriers filtering in FBMC, groups of subcarriers (sub-bands) are filtered in UFMC. Also, UFMC can still use QAM in contrast with FBMC, which enables usage of MIMO with this technology [62].

Discrete Fourier Transform-Spread-OFDM (DFTs-OFDM) is a variation of OFDM in which additional DFT step is added before OFDM modulation. It is similar to SC-FDMA. Main advantage is low PAPR (compared to OFDM) and simple equalization in the frequency domain [63].

Generalized Frequency Division Multiplexing (GFDM) is an alternative multicarrier technique that can solve the issues of using FBMC with MIMO and OFDM's long CP and usage of rectangular pulses. In GFDM, a nonorthogonal prototype filter is used to carry complex-valued symbols in time-frequency grid. To avoid IBI, a CP is added to each GFDM block. The ratio of CP length to the data block length can be smaller in GFDM compared to that in OFDM. As such, the bandwidth efficiency of GFDM can be higher than that of OFDM [61]. One method for improving the performance of GFDM to be used in networks of the future is to use it with another 6G technology Intelligent Reflecting Surfaces (IRS) [64].

Orthogonal Time Frequency Space (OTFS) waveform's main feature is that the information bearing signal is placed in the delay Doppler domain as opposed to the usual placement of similar signals in the time-frequency domain. This feature enables it to overcome several disadvantages of OFDM. OTFS is known to be more resilient to frequency offset and Doppler which is one of the key drawbacks of OFDM. OTFS can support higher mobility as well as higher frequency bands of operation. [65] It is fair to expect usage of a combination of OTFS and OFDM. For high Doppler shift scenarios, OTFS-based enhanced waveforms may be a good choice, while in indoor hotspot coverage scenarios, the advantages of OFDM multi-carrier-based enhanced waveforms ca be utilized [66].

OFDM with Index Modulation (OFDM-IM) extends OFDM's signal constellations information by adding an additional level of modulation using the subcarrier index. In the OFDM-IM system model, the subcarriers are divided into two groups - active and inactive - based on a predetermined index pattern. Indices of the subcarriers are activated according to the information bits [67].

Index Modulation Non-Orthogonal Multiple Access (IM-NOMA) is a combination of multiple access technique index modulation and NOMA. IM-NOMA is designed to allow multiple users to share the same resources, by allocating different power levels and index values to the users. It is useful in scenarios where the channel conditions of different users are similar, as it allows more users to share the same resources and increases the capacity of the system [68].

# Conclusion

In this paper we described the operation of an OFDM digital communications system chain operation (transmitter, channel and receiver) within the context of multipath fading channel. Various potential interference types are described in detail, and a survey of current interference mitigation techniques is provided with intentional interference - OFDM jamming included. A survey of potential OFDM modifications – candidates for the modulation scheme in the future 6G cellular networks is also provided.

## ◣ Acknowledgement

This work was supported by the Slovak Research Agency (https://vyskumnaagentura.sk) under the grant call Nr. 09I03-03-V04, under the application: 09I03-03-V04-00259 : project IMION - INTERFERENCE MITIGATION IN OFDM NETWORKS.

## ◣ References

▰ **Authors**

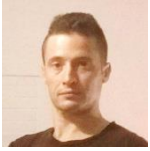

**doc. Ing. Tomáš Páleník, PhD.**
Slovak University of Technology (STU) in Bratislava, Slovakia.
tomas.palenik@stuba.sk (corresponding author)
He received his Master's degree in 2006 from the STU and in 2010 he finalized his dissertation: "Communication system design based on an SDR platform: Exploiting the redundancy of an OFDM system" and received the Ph.D. degree in telecommunications. His research interests include digital communications systems, Orthogonal Frequency Division Multiplexing, Software Defined Radio, Error-Control Coding, IPv6 & IoT and graph algorithms in communications. During 2006–2008 he worked as an external consultant for Sandbridge-Technologies, N.Y., USA, where he implemented an LDPC decoder for an in-house developed multicore mobile processor SandBlaster. Later working in academia, he has participated in multiple research projects and also took part in the COST action IC1407. In 2019 he worked as a researcher and analyst at the AIT - Austrian Institute of Technology. He presented at several international conferences such as Wireless Innovation's SDR in Washington D.C., USA.
He is a member of the IEEE.

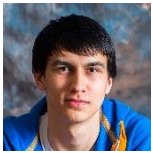

**Ing. Viktor Szitkey**
Slovak University of Technology (STU) in Bratislava, Slovakia.
viktor.szitkey@stuba.sk
Research interests focus on various technology implementation in SDR.
Student member of the IEEE.